\RequirePackage{fix-cm}
\documentclass[smallextended]{svjour3}       % onecolumn (second format)
\smartqed  % flush right qed marks, e.g. at end of proof

\usepackage[latin1]{inputenc}
\usepackage[T1]{fontenc}
\usepackage{amsmath}
\usepackage{bm}
\usepackage{bbm}
\usepackage{amsfonts}
\usepackage{amssymb}
\usepackage[usenames,dvipsnames,svgnames,table]{xcolor}
\usepackage{graphicx}
\usepackage{enumerate}
\usepackage{commath}
\usepackage{ulem}
\usepackage{hyperref}
\usepackage[numbers]{natbib}

\title{
Probability density function (PDF) models for particle transport in porous media
}
\author{Matteo Icardi \and Marco Dentz}

\institute{M. Icardi \at
              School of Mathematical Sciences, University of Nottingham, NG7 2RD, UK \\
              \email{matteo.icardi@nottingham.ac.uk}           %  \\
           \and
           M. Dentz \at
              IDAEA-CSIC
}

\date{Received: / Accepted:}

\newcommand{\Vb}{\mathbf{V}}

\newcommand{\Ub}{\mathbf{U}}
\newcommand{\ub}{\mathbf{u}}
\newcommand{\vb}{\mathbf{v}}

\newcommand{\xb}{\mathbf{x}}
\newcommand{\Xb}{\mathbf{X}}

\newcommand{\wienerb}{\mathbf{W}}
\newcommand{\wiener}{W}

\newcommand{\be}[1]{
\begin{equation}
\expandafter\label{eq:#1}
}
\newcommand{\ee}{\end{equation}}
\newcommand{\eq}[1]{Eq.~\ref{eq:#1}}

\newcommand{\bfig}[1]{
\begin{figure}
\expandafter\label{fig:#1}
}
\newcommand{\efig}{\end{equation}}

\newcommand{\f}{f}
\newcommand{\F}{F}
\renewcommand{\u}{u}
\newcommand{\x}{x}
\newcommand{\y}{y}
\renewcommand{\v}{v}
\newcommand{\X}{X}

\newcommand{\Y}{Y}

\renewcommand{\c}{x}
\newcommand{\diff}{D}
\newcommand{\deform}{\mathbf{G}}

\renewcommand{\c}{c}

\renewcommand{\O}[1]{\mathcal{O}(#1)}

\newcommand{\divx}{\nabla_{\xb}\cdot}
\newcommand{\divv}{\nabla_{\vb}\cdot}
\newcommand{\grad}{\nabla}
\newcommand{\gradx}{\nabla_{\xb}}
\newcommand{\gradv}{\nabla_{\vb}}

\newcommand{\ensavg}[1]{\left<#1\right>} % ensemble average
\newcommand{\spatavg}[1]{\overline{#1}} % spatial average
\newcommand{\frobenius}{:} % element wise sum

\newcount\colveccount
\newcommand*\colvec[1]{
        \global\colveccount#1
        \begin{pmatrix}
        \colvecnext
}
\def\colvecnext#1{
        #1
        \global\advance\colveccount-1
        \ifnum\colveccount>0
                \\
                \expandafter\colvecnext
        \else
                \end{pmatrix}
        \fi
}

\newcount\rowveccount
\newcommand*\rowvec[1]{
        \global\rowveccount#1
        \begin{pmatrix}
        \rowvecnext
}
\def\rowvecnext#1{
        #1
        \global\advance\rowveccount-1
        \ifnum\rowveccount>0
                &
                \expandafter\rowvecnext
        \else
                \end{pmatrix}
        \fi
}

\renewcommand{\matrix}[1]{\begin{pmatrix}#1\end{pmatrix}}

\begin{document}
\graphicspath{{./figures/}}

\DeclareGraphicsExtensions{.pdf,.png,.jpg}

\maketitle

\begin{abstract}
% \PACS{PACS code1 \and PACS code2 \and more}
% \subclass{MSC code1 \and MSC code2 \and more}
Mathematical models based on probability density functions (PDF) have
been extensively used in hydrology and subsurface flow problems, to describe
the uncertainty in porous media properties (e.g., permeability
modelled as random field). Recently, closer to the spirit of PDF models for
turbulent flows, some approaches have used this statistical viewpoint also in pore-scale
transport processes (fully resolved porous media models).
When a concentration field is transported, by advection and diffusion, in a heterogeneous medium, in fact,
spatial PDFs can be defined to characterise local fluctuations and improve or better
understand the closures performed by classical upscaling methods.
In the study of hydrodynamical dispersion, for example, PDE-based PDF approach can replace expensive
and noisy Lagrangian simulations (e.g. trajectories of drift-diffusion stochastic processes).
In this work we derive a
joint position-velocity Fokker-Planck equation to model the motion of particles undergoing advection and diffusion in
in deterministic or stochastic heterogeneous velocity fields.
After appropriate closure assumptions, this description can help deriving rigorously stochastic models for the statistics of Lagrangian velocities.
This is very important to be able to characterise the dispersion properties and can, for example, inform velocity evolution processes in
Continuous Time Random Walk (CTRW) dispersion models.
The closure problem that arises when averaging the Fokker Planck equation shows also interesting similarities with the mixing
problem and can be used to propose alternative closures for anomalous dispersion.
%Finally, we show a simplified approach where a joint time-position PDF can be used to extract information on the statistics of arrival times, and its moments.


\end{abstract}

%%%%%%%%%%%%%%%%%%%%%%%%%%%%%%%
%%%%%%%%%%%%%%%%%%%%%%%%%%%%%%%
\section{Introduction}
%%%%%%%%%%%%%%%%%%%%%%%%%%%%%%%
%%%%%%%%%%%%%%%%%%%%%%%%%%%%%%%

%Phenomena and motivation
The evolution of solute and particle transport in heterogeneous porous
media is determined by the heterogeneity of the medium, the consequent
flow heterogeneity and small scale diffusion~\cite{saffman1959,josselin1958,bear1972}. These processes
determine the average transport behaviours and also the fluctuation
dynamics. For example, small scale velocity fluctuations and mass
transfer processes give rise to large scale transport dynamics
characterised by hydrodynamic dispersion, but also non-Fickian
transport characteristics such as long tails in breakthrough curves
and non-linear evolution of solute
dispersion~\cite{liu2012,kang2014,deanna2013,dentz2018,puyguiraud2019b}. The
understanding of these behaviours plays a central role in a series of
applications across different fields and applications ranging from
geothermal energy to packed bead reactors. 
%Importance of Velocity PDFs
The probability density functions (PDFs) of Eulerian and Lagrangian velocities, their relation with
medium properties, and their evolution with travel distance and time
provide quantitative insight into the mechanisms of hydrodynamic
dispersion and mixing in porous media.      
  
%Review on recent efforts to characterize velocity distributions in
%porous media
The pioneering study of \citet{moroni2001} used $3$-dimensional particle tracking velocimetry
to analyze velocity PDFs in the context of a non-local theory for
hydrodynamic dispersion in porous media. The works by
\citet{deanna2013} and~\citet{kang2014} studied intermittency in purely advective
Lagrangian velocity series in $2$ and $3$-dimensional synthetic porous
media and modelled their evolution using a continuous time random walk
approach. \citet{siena2014} analysed the scaling behaviours of Eulerian
velocity statistics in $3$-dimensional digitised rocks and find
stretched exponential distributions for the stream-wise
velocities. Similar statistics were found for the distribution of
stream-wise Eulerian velocity for $2$-dimensional synthetic fibrous
media~\cite{matyka2016}. \citet{jin2016} analysed correlations in
medium and flow properties as well as distributions of stream-wise
velocities for different $3$-dimensional porous media. These authors
find similar behaviours for flow and structural correlation
functions. \citet{holzner2015} and \citet{morales2017} used particle
tracking velocimetry to study Lagrangian velocities, accelerations and
dispersion in $3$-dimensional bead packs, which are modelled using
spatial velocity Markov models. \citet{holzner2015} propose a model to
relate pore size and velocity distributions. \citet{deanna2017}
analysed Eulerian velocity distributions in $2$-dimensional synthetic
porous media and revealed a relation between the pore aperture
distribution and the distribution of the Eulerian velocity magnitude,
which is explained by Poiseuille flow in individual pores combined
with approximately constant pressure
drops~\cite{saffman1959}. \citet{meyer2016} analysed Lagrangian
velocity time series in $3$-dimensional porous media, based on which
they propose a stochastic model for the evolution of particle
velocities. \citet{alim2017} proposed a velocity model based on mass
conservation and independence of the flow rates in neighboring pore
throats to explain exponential tails for the stream-wise velocity
distribution in simple $2$-dimensional porous
media. \citet{puyguiraud2019b} conducted a thorough analysis of
Eulerian and Lagrangian flow attributes for a digitised
$3$-dimensional Berea sandstone sample with the aim of predicting and
characterising anomalous dispersion processes at the pore
scale~\cite{puyguiraud2019b}.

The above cited works focus on Eulerian and Lagrangian flow attributes
for purely advective particle motion. The classical works
by~\citet{josselin1958} and~\citet{saffman1959} considered the
stochastic motion of particles in porous media in the presence of
diffusion to derive expressions for longitudinal and transverse
hydrodynamic dispersion coefficients. \citet{saffman1959} provides an
approach for the estimation of the Eulerian velocity distribution
based in $3$-dimensional porous media based on Poiseuille flow in
individual pores and the approximation of a constant pressure drop
across different pores. It is argued that diffusion homogenises the
particle velocities within pores. \citet{dentz2018} used a similar
approach to analyse the tailing of breakthrough curves in a
$3$-dimensional synthetic porous medium within a continuous time
random approach parametrised by the Eulerian velocity distribution
combined with diffusive mass transfer. \citet{most2016} use a copula
based method to investigate advective-diffusive particle motion in
$3$-dimensional porous media. Many of the works cited above are based
on phenomenological understanding, rather than rigorously derived from
the microscopic underlying mathematical equations. While more rigorous analytical
upscaling techniques for Partial Differential Equations (PDEs), such
as homogenisation and volume averaging, can be extended to deal with
non-Fickian dispersion and non-equilibrium effects
\cite{icardibookchapter}, they lead to challenging non-local closure
problems. 

We focus here on an alternative approach based on Probability Density
Functions, in the following termed PDF methods \cite{pope2000turbulent}. PDF methods have been pioneered in
turbulence research~\cite[][]{lumley1962mathematical,Lundgren1967,port1976random,Pope1994b},
and specifically in the context of turbulent combustion~\cite[][]{pope1985pdf,fox2003computational,haworth2010progress}. In these
contexts, the PDFs of Lagrangian and Eulerian velocities and their
relations have been studied in order to systematically quantify the
statistical nature of turbulent flow. In the context of turbulent combustion,
PDF methods provide an elegant way to systematically quantify the non-linear interplay of turbulent
mixing and chemical reaction. PDF methods have been used to
determine, for example the distribution function of transported
passive and reactive scalars, and the densities of particles that are
dispersed in a turbulent flow field~\cite[][]{Pope1994b}, or by
molecular diffusion~\cite[][]{Einstein, Smoluchowski, Langevin}. 
For flow and transport processes in porous media, PDF methods have
been used for the analysis of the impact of spatial heterogeneity on
Darcy scale transport and reaction. This includes studies that map,
numerically or analytically, the statistics of random porous media flow  
 on the velocity and concentration statistics for purely advective \cite{Shvidler2003} and advective-dispersive
transport under transient and steady state conditions
\cite{nowak2008probability,sanchez2009conditional,dentz2010probability,
  Meyer2010a, Dentz2012,deBarros2014}. Other works apply PDF methods to advective-reactive transport characterized by steady random flow and
reaction conditions~\cite[][]{Lichtner2003, Tartakovsky2009,
  Tartakovsky2011, Venturi2013}. While these works derive concentration PDFs based on the
advection-dispersion or advection-reaction equation for the
transported scalar, other PDF
approaches~\cite[][]{bellin2007probability,Suciu2015a, suciu2016towards} start
from Langevin equations for the stochastic evolution of scalar
concentration, and obtain the PDF models from the corresponding
Fokker-Planck equations. PDF methods in porous media have also been used
both for the upscaling of fluctuating small scale dynamics of mixing
and reaction, and in the context of probabilistic risk assessment~\cite[][]{Tartakovsky2013}. 
 
% MD TO MATTEO: I COULD NOT FIT THE FOLLOWING INTO THE GENERAL FLOW
%
% The derivation of macroscopic (averaged) transport models, both for
% particle positions or velocities, can also be seen as a
% coarse-graining or model reduction \cite{Givon2004} of the
% high-dimensional Langevin governing equations for particle position
% and velocity. While this has been extensively studied for
% Hamiltonian-type systems
% \cite{Hijon2006,DiPasquale2019}, the coarse-graining of general
% advection-diffusion models has been only recently studied for the
% overdamped Langevin equation \cite{Duong2018,Legoll2010,Hudson2018}.

%
% These works consider purley advective transport. The   
% Velocity variability induced by porescale heterogeneity is key for the understanding of porescale transpor
% - Lagrangian velocity statistics in turbulence, difference to porous media
% - importance of diffusion as a sampling mechanism
%

In this paper we introduce an equation for the joint distribution of particle position and velocity under advection and diffusion and apply statistical mechanics techniques to obtain the equation that describes the evolution of the velocity PDF. This is done by conditional expectations and perturbation techniques. This approach has not been applied before for porous media flows although it can be regarded as a special case of coarse-graining or model reduction \cite{Givon2004} of the
 high-dimensional Langevin governing equations for particle position
 and velocity. While this has been extensively studied for
 Hamiltonian-type systems
 \cite{Hijon2006,DiPasquale2019}, the coarse-graining of general
 advection-diffusion models has been only recently studied for the
 overdamped Langevin equation \cite{Duong2018,Legoll2010,Hudson2018}.

In section 2, we derive a system of Langevin equations for the evolution of particle position and velocity and the equivalent Fokker-Planck equation for their joint probability density function (PDF). Section 3 focuses on the evolution of the marginal velocity PDFs and closure models. Section 4 provides some simplified models and exact solutions for the velocity PDF in deterministic linear shear flows. Eventually some preliminary numerical results and observations are presented in Section 5.

%%%%%%%%%%%%%%%%%%%%%%%%%%%%%%%
%%%%%%%%%%%%%%%%%%%%%%%%%%%%%%%
\section{Joint position-velocity PDF equation}
%%%%%%%%%%%%%%%%%%%%%%%%%%%%%%%
%%%%%%%%%%%%%%%%%%%%%%%%%%%%%%%

Scalar and particle (advective and diffusive) transport can be modelled with the Ito (also known as overdamped Langevin) Stochastic Differential Equation (SDE)
\be{sde_pos}
\dif \Xb= \ub(\Xb)\dif t + \sqrt{2\diff}\dif \wienerb
\ee
with $\Xb(t=0)=\Xb_{0}$, where $\wienerb$ is a $n$-dimensional Brownian motion, $D$ a diffusion constant, and $\ub(\xb)$ is a space-dependent constant velocity field. If $\Xb, \Vb, \ub$ are considered dimensionless, it is enough to replace $D$ with the inverse P\'eclet number, $\mathrm{Pe}$. Equivalently, this can be described by the Fokker-Planck equation
\be{fp_pos}
\dpd{\c}{t}=-\divx\del{\ub(\xb)\c}+\Delta_\xb\del{\diff\c}
\ee
where $\Delta$ is the Laplacian operator, with initial conditions $\c(\xb;t=0)=\delta(\xb-\Xb_{0})$.
The probability density function (PDF) $\c=\c(\xb;t)$ can be also interpreted as a concentration field as it is obtained by the  SDE \eq{sde_pos} as
\be{fp_def}
\c=\ensavg{\delta(\Xb-\xb)}
\ee
where the operator $\ensavg{\cdot}$ indicates the average with respect to the Brownian motion (stochastic average).

The problem with this formulation is that we lose completely the information on the trajectories and, particularly, on the Lagrangian velocities $\Ub=\ub(\Xb)$ and their time evolution. One way to restore it is to introduce a joint position-velocity PDF.
We can extend the Markovian variable to the joint variable $(\Xb,\Ub)$ and write a system of Ito SDEs, by writing $\Ub$ using Ito's lemma and remembering that we have a steady flow field so that the time derivative disappears. Equivalently one can think of it as a Taylor expansion around the point $\Xb$
\be{du_exp}
\dif \Ub=\dif\ub(\Xb)=\gradx\ub\dif \Xb+\frac{1}{2}\dif \Xb^{T}\gradx^{2}\ub\dif \Xb+\O{\dif \Xb^{3}}
\ee
where $\ub=\ub(\Xb)$ is the velocity field evaluated in the position
$\Xb$ and, and $\grad^{2}$ indicating the Hessian operator. Note that
Ito's lemma requires an expansion of $\dif \Ub$ up to second order in
$\dif \Xb$ in order to obtain a consistent expansion in $\dif t$
because the random increment $\dif \wienerb$ is of order $\dif t^{1/2}$. 
Substituting \eq{sde_pos} in \eq{du_exp} we obtain an explicit SDE for the Lagrangian position and velocity. Considering only the first order terms and remembering that $\dif \wiener^{2}=\dif t$,  
\be{sde_vel}
\dif\colvec{2}{\Xb}{\Ub}=\colvec{2}{\ub}{\deform\Ub+\diff\Delta\ub}\dif t + \sqrt{2\diff}\colvec{2}{\mathcal{I}_{n\times n}}{\deform}\dif\wienerb 
\ee
where $\deform=\deform(\Xb)=\eval[1]{\grad\ub}_{\Xb}$ is the deformation tensor evaluated at $\Xb$ and $\mathcal{I}_{n\times n}$ is the identity matrix of order $n$ (spacial dimensions and length of vectors $\Ub$ and $\Xb$). If we assume that $\ub$ is the solution of a stationary Stokes flow, then the term $\Delta{\ub}=\frac{1}{\mu}\grad{p}$, where $p$ is the pressure.

The corresponding Fokker-Planck equation \cite{Kampen2007,Gardiner2009} for the joint PDF $\f(\xb,\vb;t)$ can be therefore written as\footnote{
We are dealing with a $2n$-dimensional Fokker-Planck equation whose $2n\times2n$ diffusion matrix is
$$
\frac{1}{2}\colvec{2}{\sqrt{2\diff}\mathcal{I}_{n\times n}}{\sqrt{2\diff}\deform}
\colvec{2}{\sqrt{2\diff}\mathcal{I}_{n\times n}}{\sqrt{2\diff}\deform}^{T}
=\diff
\matrix{\mathcal{I}_{n\times n} & \deform^{T}\\ \deform & \deform\deform^{T}}
$$
}
\be{fp_joint}
\dpd{\f}{t}= -\divx\left[\vb\f\right] +\Delta_{\xb}\left[\diff\f\right]  % space
- \divv\left[(\deform\vb + \diff\Delta\ub)\f\right] % velocity advection
+\gradv^{2}\frobenius\left[\diff(\deform\deform^{T})\f\right]  % diffusion
+ 2\gradx\gradv\frobenius\left[\diff\deform\f\right] % mixed terms
\ee
where we have denoted the Frobenius double inner product with ``$\frobenius$'' notation.

\eq{fp_joint} can  also be written in a fully conservative way, as follows:
\be{fp_joint2}
\dpd{\f}{t}= \divx\left[(-\vb+\diff\gradx)\f\right]  % space
+ \divv\left[(-\deform\vb - \diff\Delta\ub+\diff\deform\deform^{T}\gradv)\f\right] % velocity
%+ 2\divx\left[\diff\deform\gradv\f\right] % mixed terms
\ee
where the change of sign of the term $\diff\Delta\ub$ is due to a cancellation with the integration by parts of the mixed derivative. Another useful formulation is obtain by bringing outside derivatives all terms:
\be{fp_joint3}
\dpd{\f}{t}+ \vb\cdot\gradx\f
+ (\deform\vb + \diff\Delta\ub)\cdot\gradv\f
=
\diff\Delta_{\xb}\f+
\diff(\deform\deform^{T})\frobenius\gradv^{2}\f
%+ 2\diff\deform\frobenius\gradx\gradv\f
\ee

In the equivalence of these three forms of the joint position-velocity PDF, we have assumed here
 non-inertial particles\footnote{This means that the advective velocity is always equal to the flow velocity and depends only on the spatial coordinates. In the case of inertial particles this it not true anymore since one can define a advective velocity that is a function of the fluid velocity. This will be studied in our future work.} (i.e., $\gradv\ub=0$ such that $\gradv\deform=0$) and an incompressible flow field ($\divx\ub=0$)\footnote{ In fact, when $\deform\vb$ is taken out of the velocity derivative
$$
\divv(\deform\vb\f) = \mbox{tr}(\deform)\f + (\deform\vb)\cdot\gradv\f
$$
and the first is null only for incompressible fluids.
}.

These equations represent a novel point of view in the analysis of transport and mixing in heterogeneous media. While the advection and diffusion in space might look similar to the standard advection-diffusion equation, their coefficients are no longer explicitly depending on space but only on the now internal coordinate $\vb$. This means that the advection term $\divx\left[\vb\f\right]$ can now be integrated and/or averaged in space with no closure. This is at the expenses of introducing three additional dimensions for velocity. As we will discuss in the next section, if the equation is averaged spatially, the closure problems arise instead from the terms that contain derivative in the velocity space. 

Let us focus our analysis here to \eq{fp_joint2}. Two different velocity-drift (advection) terms (i.e. acceleration/deceleration) can be identified, and the diffusion term in the velocity. The latter, compared to the standard diffusion in space, has now a non-constant coefficient that depends on the local shear rate tensor.
The diffusivity also enters into the second velocity-drift advection term due to the diffusion-induced changes of velocities. This can be explained by considering the effect that spatial diffusion  has on the velocity. For example, if a particle lies in the slowest (wrt fastest) regions of the flow, the diffusive jumps will cause a  change in the velocity that is not isotropic (like for changes of position in spatial diffusion) but instead weighted by the local shear and will depend on the local velocity, causing therefore  an additional velocity-drift term (similarly to a non-constant diffusion in space that cause an additional drift term).

%%%%%%%%%%%%%%%%%%%%%%%%%%%%%%%
%%%%%%%%%%%%%%%%%%%%%%%%%%%%%%%
\section{Velocity marginal PDF and closures\label{sec:marginal}}
%%%%%%%%%%%%%%%%%%%%%%%%%%%%%%%
%%%%%%%%%%%%%%%%%%%%%%%%%%%%%%%

In certain cases, we can be interested solely on the evolution of the velocity PDF. We can therefore marginalise the general  \eq{fp_joint}, by defining $\F=\int\f\,\dif\xb=\spatavg{\f}$ as the global spatial average\footnote{This is not to be confused with the stochastic average used above. To distinguish, the  spatial average  is denoted by $\spatavg{\cdot}$, while the stochastic average with respect to the Brownian/dispersive motion will be denoted by  $\ensavg{\cdot}$} and neglecting all the boundary terms arising from spatial derivatives (assuming an infinite or periodic domain), obtaining
\be{fp_joint_marg}
\dpd{{\F}}{t}=-\int\divv[(\deform\vb-\diff\Delta\ub)\f]\,\dif\xb+\int\diff(\deform\deform^{T})\frobenius\gradv^{2}\f\,\dif\xb
\ee
A similar result would be obtained by local spatial averages (filtering) with the difference that the spatial derivative will not disappear. In this work, we want to focus on the closure of the terms that depend on derivatives in the velocity space, and in the dynamics in the velocity space. Local spatial averages will be the subject of further  investigations.

\eq{fp_joint_marg} is now an equation in $3+1$ dimensions (velocity and time), that represent the evolution of Lagrangian particle velocities in the whole space. This is however unclosed as the last two terms contain non-trivial spatial integrals of velocity gradients ($\deform$ and $\Delta\ub$) weighted by the joint pdf $\f$.
To approximate these integrals, we can decompose $\f(\xb,\vb;t)=\f(\xb|\vb;t)\F(\vb;t)$ and, denoting conditional averaging over $\f(\xb|\vb;t)$ with superscript $\xb|\vb$, the equation can be written as

\be{fp_joint_marg_cond_clos}
\dpd{{\F}}{t}=-\divv\left[\left(\spatavg{\deform}^{\xb|\vb}\vb-\diff\spatavg{\Delta \ub}^{\xb|\vb}\right)\F\right]+\Delta_\vb\left(\diff\spatavg{\deform\deform^{T}}^{\xb|\vb}\F\right)
\ee
In general, compared to the simplified case considered in
Sec.~\ref{sec:shear}, the flow map that links each position to a
velocity might not be invertible, making the conditional expectations
not trivial to compute. Furthermore, the conditional expectations in general depend on time. To better understand the physical meaning of these conditional expectations,  we can rewrite the conditional PDF (see also an alternative derivation in Appendix A), using Bayes formula, as:
\be{bayes}
\f(\xb|\vb;t)
=
\frac{\f(\vb|\xb)\f(\xb;t)}{\F(\vb;t)}
=
\frac{\delta(\vb-\ub(\xb))\c(\xb;t)}{\F(\vb;t)}
\ee
where the marginal PDF of position $\f(\xb;t)=\c(\xb;t)$ is the concentration field, and the conditional probability $\f(\vb|\xb)$ is a deterministic function that maps each position into its velocity\footnote{Since our model assume that the particles instantaneously relax to the local, constant in time, fluid velocity field.}, i.e. it is the delta function  $\delta(\vb-\ub(\xb))$.
The appearance of the delta function inside the conditional probability results in the closure integral terms to be evaluated only on the domain $\Omega_{\vb}=\{\xb\; | \; \ub(\xb)=\vb\}$ with the concentration field as a weight, i.e., for a generic function $g=g(\xb)$,
\[
\spatavg{g(\xb)}^{\xb|\vb}=\int_{\Omega_{\vb}}g(\xb)\c(\xb;t)\dif\xb
\]

As we will see in Sec.~\ref{sec:constant_shear}, this closure gives an exact analytical formula for simple shear flows. More generally, in \eq{fp_joint_marg_cond_clos}, one has to compute these closure terms as functions of velocity $\vb$ and time $t$, or relying on equilibrium approximations (see Sec.~\ref{sec:equil}), in which the invariant (long-time) concentration (position PDF) $\c_e$ is used instead, and the corresponding marginal velocity PDF at equilibrium $\F_e(\vb)$, defined as:
\be{Feq_def}
\F_e(\vb) = \lim_{t\to\infty} \F(\vb;t)
\ee

%%%%%%%%%%%%%%%%%%%%%%%%%%%%%%%
\subsection{
Constant shear}\label{sec:constant_shear}
%%%%%%%%%%%%%%%%%%%%%%%%%%%%%%%
For the simple case of constant shear rate, $\deform$ is a constant matrix, and \eq{fp_joint_marg_cond_clos} can be explicitly closed and written as an advection diffusion reaction equation
\be{fp_joint_marg_constG}
\dpd{{\F}}{t}=-(\deform\vb)\cdot\gradv\F - \mbox{tr}(\deform) \F +\diff(\deform\deform^{T})\frobenius\gradv^{2}\F
\ee
where we have taken $(\deform\vb)$ out of the derivative and $\mbox{tr}(\deform)=0$ for an incompressible fluid. The stationary (equilibrium) velocity PDF $\F_e(\vb)$, can be calculated by removing the time derivative. This can be formally integrated (similarly to the stationary distribution of the
Ornstein--Uhlenbeck process), provided that $\deform$ is an invertible matrix, resulting in a Gaussian solution 
\[
\F_e(\vb)\propto \exp^{-\vb^T\deform^T\deform^{-1}\deform\vb}
\]
This is, however, a well-defined probability only for positive definite matrix $\deform$, which is in contradiction with assumption of it being traceless. A physical explanation of this is that an incompressible fluid in an infinite domain with a constant shear necessarily brings particles towards region of infinite velocity. If we drop instead the assumption of incompressibility and we assume a positive definite $\deform$, the particles will all be driven into the origin and oscillate around it by diffusion, assuming therefore a Gaussian distribution of velocities.

%%%%%%%%%%%%%%%%%%%%%%%%%%%%%%%
\subsection{Equilibrium Eulerian velocity PDF}\label{sec:equil}
%%%%%%%%%%%%%%%%%%%%%%%%%%%%%%%
At equilibrium, the concentration $\c(\xb,t)$, for solenoidal (divergence-free) velocity fields $\ub$, tends to a constant value\footnote{Provided appropriate boundary conditions, such as local periodicity. For more general cases, with reactions, sources or sinks, or with non-solenoidal velocity fields, the equilibrium concentration can be a non-constant function.} $\c_e$. Therefore, the joint PDF at equilibrium is constant in $\xb$ and simply proportional to $\F_e(\vb)$, and the conditional expectations (spatial averages against the conditional distribution), for a generic function $g$, are defined as
\be{eq_cond}
\spatavg{g(\xb)}^{e|\vb}
=
\int_{\Omega_{\vb}}g(\xb)\f_e(\xb|\vb)\dif\xb
=
\int_{\Omega_{\vb}}g(\xb)\frac{\delta(\vb-\ub(\xb))\c_e}{\F_e(\vb)}\dif\xb
\ee
where, as above, we have used Bayes formula to express the equilibrium conditional probability. As it can be seen, this expectation is now independent of time, as long as the velocity field does not depend on time.
%Therefore, the term $\spatavg{\deform}^{\xb|\vb}=\spatavg{\deform}$ becomes simply a term dependent on the stretching at the boundary, while $\spatavg{\Delta \ub}^{\xb|\vb}=\frac{1}{\mu}\spatavg{\grad{p}}$ depends only on the average pressure gradient at the boundaries, if we assume the flow is given by Stokes flow.

The stationary version of \eq{fp_joint_marg_cond_clos}, i.e., 
\be{Feq}
\divv\left[\left(\spatavg{\deform}^{e|\vb}\vb-\diff\spatavg{\Delta \ub}^{e|\vb}\right)\F_e\right]
=
\Delta_\vb\left(\diff\spatavg{\deform\deform^{T}}^{e|\vb}\F_e\right)
\ee
is therefore the equation satisfied by the equilibrium velocity PDF $\F_e(\vb)$ which, in this case, it is equivalent to the Eulerian velocity PDF, i.e. the velocity PDF obtained by sampling uniformly\footnote{Since we have here assumed a constant equilibrium concentration.} the whole domain.

%%%%%%%%%%%%%%%%%%%%%%%%%%%%%%%
\subsection{Perturbation near equilibrium}
%%%%%%%%%%%%%%%%%%%%%%%%%%%%%%%
To obtain a simpler dynamics, we can now decompose the velocity PDF near the equilibrium as:
\[
\F(\vb,t)=\F_e(\vb)+\F^*(\vb,t)
\]
and decompose also the conditional expectations as:
\[
{\f}\del{\xb,t|\vb} = \f_e(\xb|\vb)  + {\f}^{*}\del{\xb,t|\vb}
\qquad \Longrightarrow \qquad
{g}^{\xb|\vb} = {g}^{e|\vb} + {g}^{*|\vb}
\]
where the latter, being the difference between two conditional distributions, is not a distribution (e.g., it does not integrate to one but instead its integral is null) and it is not directly computable from fluctuations $\c(\xb,t)-\c_e$, as the conditional distributions in \eq{bayes} and \eq{eq_cond} have a different denominator.

Applying these decompositions to \eq{fp_joint_marg_cond_clos}, and using the definition of $\F_e$, \eq{Feq},  we obtain:
\be{Fprime}
\dpd{{\F^*}}{t}+\divv\left[\left(\spatavg{\deform}^{e|\vb}\vb-\diff\spatavg{\Delta \ub}^{e|\vb}\right)\F^*\right]
-
\Delta_\vb\left(\diff\spatavg{\deform\deform^{T}}^{e|\vb}\F^*\right)
=
-\tau(\vb,t,\F^*)
\ee
where
\be{Fprimetau}
\tau(t,\vb,\F^*)=\divv\left[\left(\spatavg{\deform}^{*|\vb}\vb-\diff\spatavg{\Delta \ub}^{*|\vb}\right)\F^*\right]
-
\Delta_\vb\left(\diff\spatavg{\deform\deform^{T}}^{*|\vb}\F^*\right)
\ee
where we have used the fact that $\tau(t,\vb,\F^e)=0$ to preserve the correct equilibrium.

\eq{Fprime} is now an equation for $\F^*$ with coefficients constant in time. They have however a possibly complex dependence on the independent variable $\vb$, through the conditional expectation. Furthermore, the forcing term $\tau$ is still a complex time-dependent term that still needs a closure.

Since $\spatavg{{\f}^{*}\del{\xb|\vb}}=0$, the easiest closure, used below in the numerical example in Sec.~\ref{sec:shear}, is to assume the terms to be averaged are uncorrelated with $\f^*$, resulting in $\tau=0$. Alternatively, another possibility that will be considered in future works, is to lump this term into  an relaxation term to drive the system towards equilibrium, i.e., $\tau(t,\vb,\F^*)\approx \tau_0(\vb)F^*$.

\subsection{Interaction by Exchange with the Mean}
%%%%%%%%%%%%%%%%%%%%%%%%%%%%%%%
We introduce here two simplifying assumptions that allows to rewrite the parameters given by the conditional expectations in \eq{Fprime}, that still depend in a complex way on the independent variable $\vb$, in a simpler form. Similarly to the IEM model (Interaction by Exchange with the Mean) in turbulent mixing \cite{pope2000turbulent}, and to the Mori Projector in Statistical Mechanics \cite{grabert2006projection}, by assuming that velocity $\vb$ and velocity gradient $\deform$ are spatially distributed according to a joint Gaussian distribution, or, equivalently, approximating $\deform$ as a linear function of $\vb$, we can rewrite\footnote{using the formula for conditional expectations of joint Gaussian distributions
\[
\mathbb{E}[Y|X] =  \mathbb{E} [Y] + \frac{\mathrm{cov}(X, Y)}{\mathrm{var}(X)} (X - \mathbb{E}[X])
\]
or, equivalently, interpreting the conditional expectation as a projection into the reduced space span by linear functions of $\vb$}
the first two conditional expectations in \eq{Fprime} as follows:
\be{iem}
\spatavg{\deform}^{e|\vb}
\approx
\left(\vb-\spatavg{\ub}\right)^{T}\spatavg{\ub'\ub^{'T}}^{-1}\spatavg{\ub'\otimes\deform}
\ee
\be{iem2}
\spatavg{\Delta \ub}^{e|\vb}
\approx
-\left(\vb-\spatavg{\ub}\right)^{T}{\spatavg{\ub'\ub^{'T}}^{-1}}{\spatavg{\deform\deform^{T}}}
\ee
where we have assumed that $\spatavg{\deform}=0$ and $\ub'={\ub-\spatavg{\ub}}$.
All averages are now unconditional averages  and can be taken out of the derivatives. 
These averages are now more easily computable from the Eulerian velocity field.
$\spatavg{\ub'\ub^{'T}}$ is, in fact, the spatial correlation matrix of the velocity field and  $\spatavg{\deform\deform^{T}}$, that appears with an opposite sign due to the integration by parts, is a contraction of the full velocity gradient covariance $\spatavg{\deform\otimes\deform}$.

One could apply the same closure also for the third expectation in \eq{Fprime} related to the second order velocity derivative. However, here, assuming that $\deform\deform^T$ is a linear function of $\vb$ (or equivalently a joint Gaussian) is not consistent with the same assumption for $\deform$. Also, applying this closure, we can end up with possibly negative diffusion coefficients in the velocity space.
%\be{iem3}
%\spatavg{\deform\deform^{T}}^{e|\vb}
%%= \spatavg{\diff(\deform\deform^{T})}\frobenius\gradv^{2}\F + \int\dif\xb\,\diff(\deform\deform^{T})'\frobenius\gradv^{2}(\f) \\
%\approx
%%\spatavg{\diff(\deform\deform^{T})}\frobenius\gradv^{2}\F
%\left(\vb-\spatavg{\ub}\right)^{T}{\spatavg{\ub'\ub^{'T}}^{-1}}{\spatavg{\ub'\otimes\deform\deform^{T}}}
%\ee

%The last term $\spatavg{\ub\otimes\deform\deform^{T}}$, despite looking more complicated, can be equally computed as a simple space integration, given the velocity field and its derivatives.

This IEM approximation could have been equally applied directly to \eq{fp_joint_marg_cond_clos} but, without the equilibrium approximation, all the averages would still depend on time through the concentration that would act as a weighting function in the spatial averaging. This means that the evolution equation for the velocity PDF would noto be fully closed since it would require the solution of the concentration field at all times.

%%%%%%%%%%%%%%%%%%%%%%%%%%%%%%%%
%\subsection{Diagonal approximation}
%%%%%%%%%%%%%%%%%%%%%%%%%%%%%%%%
%In the simple case of all correlation tensors being diagonal, one can try to isolate advection, diffusion and reaction terms appearing in the equation
%
%TBD

%%%%%%%%%%%%%%%%%%%%%%%%%%%%%%%%
%\subsection{Limit of zero diffusion:}
%%%%%%%%%%%%%%%%%%%%%%%%%%%%%%%%
%
%For non-diffusive transport (i.e. $\diff=0$) the equation is simplified as
%\be{fp_joint2}
%\dpd{\f}{t}=-\vb\cdot\gradx\f-(\deform\vb)\cdot\gradv\f
%\ee
%\red{...continue...}
%
%
%

%%%%%%%%%%%%%%%%%%%%%%%%%%%%%%%
%%%%%%%%%%%%%%%%%%%%%%%%%%%%%%%
\section{Deterministic shear flows}
%%%%%%%%%%%%%%%%%%%%%%%%%%%%%%%
%%%%%%%%%%%%%%%%%%%%%%%%%%%%%%%

Consider the following 2D shear-flow dynamics
\be{2dshear}
\dif \colvec{2}{\X}{\Y}=\colvec{2}{\u(\y)}{\v}\dif t + \matrix{\sqrt{2\diff_{x}} & 0\\ 0 & \sqrt{2\diff_{y}}}\dif\wienerb
\ee
where $\u$ is a shear flow depending on $\y$ only (and possibly in a random manner), $\v$ is a generic function (specified later), and $\wienerb$ is a two-dimensional Wiener process. In the following we will consider a few special cases of this dynamics.

Let us assume $\diff_{x}=0$, i.e., molecular diffusion is acting only on $y$.
The joint position-velocity PDF can be written for $\f=\f(\x,\y,\u,\v;t)$ where the dependence on the velocity field is only through the velocity gradient\footnote{We assume the velocity field is smooth enough, i.e. with a Gaussian correlation. In this way the derivative exists and it is Gaussian itself with Gaussian correlation function.} $\sigma=\dpd{\u}{\y}$ and, from now on, $u$ represents only the internal velocity coordinate.

Under these assumptions,
$$
\deform=\matrix{0&\sigma\\ 0 & 0}\qquad \deform\deform^{T}=\matrix{\sigma^{2}&0\\ 0 & 0}
$$
and a few simplifications can be made on the general Fokker-Planck \eq{fp_joint2}, leading to the following equation:
\be{2dshear4_pdf}
\dpd{\f}{t}=-\dpd{}{\x}\left(\u\f\right)
-\dpd{}{\u}(\sigma\v+D\sigma')\f
+\diff\dpd[2]{\f}{\y}
+\diff\dpd[2]{}{\u}\left(\sigma^{2}\f\right)
%+2\diff\dpd{}{\x}\dpd{\f}{\v}
\ee
where  $\sigma'=\dpd{\sigma}{\y}$.
Since no significant dynamics happens in the $y-$ component of the velocity, $\v$, if we start from an initial condition $\f_{0}=\delta(\v)g(\u,\x,\y)$, the term  $\sigma\v\dpd{\f}{\u}$ can be disregarded.

As done for the general case (see Sec.~\ref{sec:marginal}), we now perform the marginalisation to obtain an equation for the velocity PDF $\F(u,t)$.  In an infinite channel, the derivatives in $x$ disappear, while integrating along the channel height $y$, the second-order derivative in $y$ can be written as a diffusive flux at the boundary walls. Since no particles can enter or exit through the walls, they are null and we obtain:
\be{2dshear_marginal}
\dpd{\F}{t}=
-\diff \dpd{}{\u}\left(\mathcal{M}_{\sigma'}\F\right)
+\diff\dpd[2]{}{\u}\left(\mathcal{M}_{\sigma^2}\F\right)
\ee
where
$\mathcal{M}_{\sigma'}(u;t)=\int\sigma'\f(y|u;t)\dif y$
and
$\mathcal{M}_{\sigma^2}(u;t)=\int\sigma^2\f(y|u;t)\dif y$.
As shown in Sec.~\ref{sec:marginal}, this depends, in general on time, as $\f(\y|\u;t)\propto\delta(y-y(u))c(y;t)$. However, as expected in a stratified flow, the evolution of velocities is due solely to the diffusion.
%Defining a P\'eclet number as $\mathrm{Pe}=\frac{u_0 L}{\diff}$, and $\tilde{t}=\mathrm{Pe}\,t$, we obtain:
%\be{2dshear_marginal_adim}
%\dpd{\F}{\tilde{t}}=
%- \dpd{}{\u}\left(\mathcal{M}_{\tilde{\sigma}'}\F\right)
%+\dpd[2]{}{\u}\left(\mathcal{M}_{\tilde{\sigma}^2}\F\right)
%\ee

%%%%%%%%%%%%%%%%%%%%%%%%%%%%%%%
\subsection{Constant shear (Couette flow)}
%%%%%%%%%%%%%%%%%%%%%%%%%%%%%%%
Couette flow is characterised by $\sigma=\sigma_0$. The velocity marginal PDF does not need any closure. This reduces to a simple heat equation for $\F(u)$.
\be{2dshear_constant_pdf}
\dpd{\F}{t}=\diff\sigma^{2}\dpd[2]{\F}{\u}
\ee
As expected, starting from any initial condition, the particles will reach a final equilibrium velocity distribution which is a linear function between the two values of velocity at the boundaries.
At early times, starting from a delta-distributed concentration of particles, the velocity PDF will evolve as a Gaussian distribution before feeling the presence of the walls.

%%%%%%%%%%%%%%%%%%%%%%%%%%%%%%%
\subsection{Linear shear (Hagen-Poiseuille flow)}
\label{sec:shear}
%%%%%%%%%%%%%%%%%%%%%%%%%%%%%%%
%
This is the case of a two-dimensional channel flow with $\sigma=-2y\frac{u_{0}}{L^{2}}$ and $\sigma'=-2\frac{u_{0}}{L^{2}}$. 
In this case, there exists a deterministic relation between $y$ and $u$, i.e.,  $u(y)=\frac{u_{0}}{L^{2}}(L-y)(L+y)=u_{0}(1-\frac{y^{2}}{L^{2}})$, such that
 %$M_{2}^{y|u}(u)=L^{2}\left(1-\frac{u}{u_{0}}\right)$, since
$
y^{2}=L^{2}\left(1-\frac{u}{u_{0}}\right)$ and 
the solution for long time limits, near equilibrium, can be computed analytically.

% \qquad \dpd{y}{u}=-L\left(1-\frac{u}{u_{0}}\right)^{-1/2}
%\qquad \dpd{u}{y}=-2y\frac{u_{0}}{L^{2}}
%$$

%\be{2dshear_linear_pdf}
%\dpd{\F}{t} = - \dpd{}{\u}\int_{y}\frac{2\diff u_{0}}{L^{2}}\f +
%\dpd[2]{}{\u}\int_{y}\frac{4\diff u_{0}^{2}}{L^{4}}y^{2}\f 
%\ee
%If we write $\f=\f(y|u)\F(u)$, the last term can be written as
%\be{2dshear_linear_pdf_cond}
%\dpd[2]{}{\u}\int_{y}\frac{4\diff u_{0}^{2}}{L^{4}}y^{2}\f=\dpd[2]{}{\u}\left(\frac{4\diff u_{0}^{2}}{L^{4}}M_{2}^{y|u}(t) \F\right)
%\ee
%where $M_{2}^{y|u}(u,t)=\int_{y}y^{2}\f(\y|\u;t)$ is the conditional second order moment.

\paragraph{Equilibrium closure:}
With the equilibrium closure, considering a constant concentration $c(x,y)=c_e$, $\mathcal{M}_{\sigma'}$ and $\mathcal{M}_{\sigma^2}$ are constant in time.
\eq{2dshear_marginal} can be therefore closed as an inhomogeneous diffusion in the velocity space:
\be{2dshear_linear_pdf2}
\dpd{\F}{t}=
\frac{4\diff u_{0}^{2}}{L^{2}} \dpd[2]{}{\u}\left(\left(1-\frac{u}{u_{0}}\right)  \F\right)
+ \frac{2\diff u_{0}}{L^{2}}\dpd{\F}{\u}=
\frac{2\diff u_{0}}{L^{2}} \dpd{}{\u}\left(2\left(u_{0}-{u}\right)  \dpd{\F}{u}- \F\right)
\ee
The stationary solution of this equation is given by:
\be{2dshear_linear_longtime}
\F \propto \left(1-\frac{\u}{\u_{0}}\right)^{-1/2}
\ee
This solution is, in fact, consistent with the fact that, when particles are uniformly spread throughout the channel, their velocity is simply given by the inverse function of the velocity profile.

\begin{figure}[htb!]
\begin{center}
\includegraphics[width=.8\textwidth]{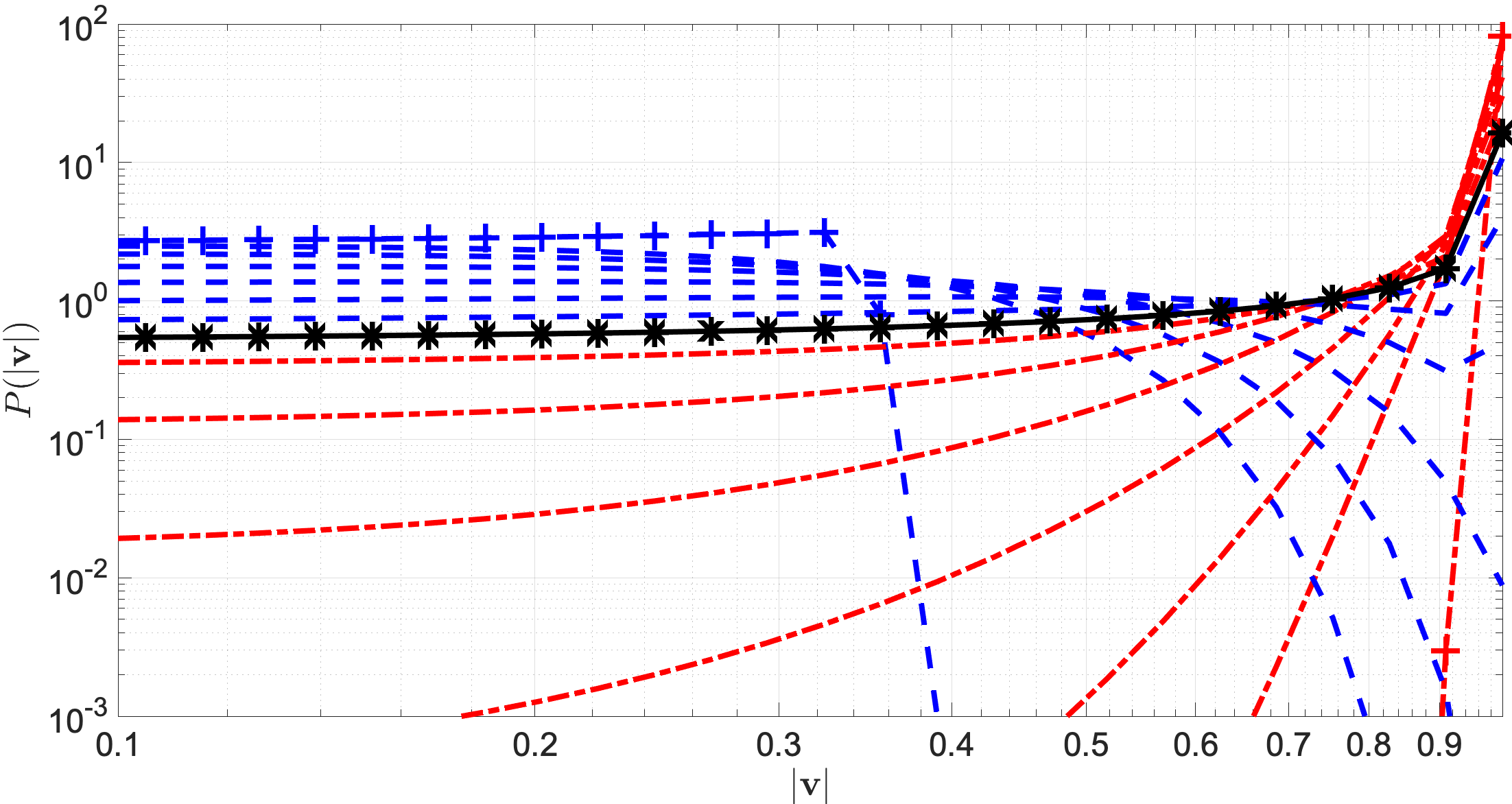}
\end{center}
\caption{Evolution of the Lagrangian velocity PDF for Poiseuille flow, starting from particles concentrated in  fast (red dot-dashed curves) or  slow (blue dashed curve) regions. The initial conditions are recognisable by added cross symbols. The equilibrium (Eulerian velocity) distribution is represented by the black solid line with symbols). The different curves show the evolution for dimensionless (advective-based) times $t=0.005; 0.1; 0.2; 0.4; 0.8; 1.6$}
\label{fig:PDFshear}
\end{figure}

In Fig.~\ref{fig:PDFshear}, we test the equilibrium closure and its convergence to the Eulerian velocity PDF. \eq{2dshear_linear_pdf2} is solved with the Matlab library Chebfun\cite{driscoll2014chebfun}, for $L=1,\,u_0=1,\,D=1$, on the domain $[0,1]$, with no-flux boundary conditions. The initial conditions (crosses) are such that particles (or solutes) are injected only in 20\% highest (red dot-dashed, wrt 20\% lowest, blue dashed) velocity regions, and the evolution towards the Eulerian velocity distribution (black continuous curve with asterisks) is depicted for different times ($t=0.005; 0.1; 0.2; 0.4; 0.8; 1.6$). As it can be seen, both the initial conditions converges relatively fast towards the equilibrium distribution. In this equation only diffusion is the driving force for this relaxation, therefore, rescaling appropriately the time, the solution is independent of the P\'eclet number.
\section{Discussion and conclusions}
%%%%%%%%%%%%%%%%%%%%%%%%%%%%%%%
%%%%%%%%%%%%%%%%%%%%%%%%%%%%%%%
In this paper, we have derived a joint position-velocity Probability Density Function (PDF) model for passive particles flowing in a porous medium or a heterogeneous flow field. This is achieved by extending the phase space and writing down the Stochastic Differential Equation (SDE) and the corresponding Fokker-Planck Partial Differential Equation (PDE). While the (Smoluchowki) position-only PDF equation contains classical advection terms $\ub\gradx{c}$, this new formulation requires only first and second-order derivatives of the flow field that often have higher-frequency oscillations and therefore, better mixing/averaging properties. This aspect, possibly leading to more accurate macroscopic models, will be explored in our future works.

Another advantage of this joint position-velocity formulation is that it can be easily extended to full second-order models with velocity increments given, for example, by the drag force as follows:
\be{sde_vel2}
\dif\colvec{2}{\Xb}{\Ub}=\colvec{2}{\ub}{\frac{\Ub-\ub(\Xb)}{St}}\dif t + \sqrt{2\diff}\colvec{2}{0}{\mathcal{I}_{n\times n}}\dif\wienerb 
\ee
Here $St$ is the Stokes number, that depends on the particle size and density, and the slip velocity (difference between particle velocity and local fluid velocity) is $\Ub-\ub(\Xb)$). In this alternative model, it is more appropriate to put the random fluctuating term in the momentum, as in the classical Langevin dynamics. %\red{What's the difference when $St<<1$? when solving for the marginal or joint velocity PDF?}
This problem is more complex as the particle velocities are no longer divergence free. This case will be studied in future works.

Since the joint PDF model is defined on a high-dimensional space (6+1 dimensions), we have studied  reduced equations obtained by integrating (marginalising) over space (assuming therefore periodicity or ergodicity) to focus on the evolution of Lagrangian particle velocities. Connections with the theory of coarse-graining are established and several possible closures are proposed. This approach, first derived for a general flow field, is then applied for two-dimensional shear flows. Here, the effect of the equilibrium, and short-time approximations are analysed.

The study of the evolution of particle Lagrangian velocities is motivated by recent studies that highlight the importance of modelling the evolution of Lagrangian velocities \cite{dentz2016continuous,dentz2018mechanisms} for predicting anomalous transport in heterogeneous media. For the first time, in this work, we propose a rigorous approach to derive an evolution equation for the Lagrangian velocity PDF.

Future works will include the application of this theoretical study to three-dimensional periodic and random porous media will be performed, the derivation of local averaged joint pdf equations (averaged in a small spatial and velocity representative volume), as well as the consequences of the closures proposed for developing improved Lagrangian and Eulerian non-Fickian transport models.

\begin{acknowledgements}
%Funding: marco's ERC, Nottingham REF grant
This work has been funded by the European Union's Horizon 2020 research and innovation programme, grant agreement number 764531, "SECURe - Subsurface Evaluation of Carbon capture and storage and Unconventional risks".
The support of the HPC Midlands Plus and Nottingham Supercomputing facilities is gratefully acknowledged.
\end{acknowledgements}

\clearpage
\bibliographystyle{plainnat}
\bibliography{joint_pdf}

\appendix

%%%%%%%%%%%%%%%%%%%%%%%%%%%%%%%
%%%%%%%%%%%%%%%%%%%%%%%%%%%%%%%
\section{Alternative derivation of $\F$ and $\f$}
%%%%%%%%%%%%%%%%%%%%%%%%%%%%%%%
%%%%%%%%%%%%%%%%%%%%%%%%%%%%%%%
Starting again from \eq{du_exp}, one can also formally write $\f$
\be{joint_pdf_def}
\begin{split}
\f(\xb,\vb,t)&=\ensavg{\delta\left(\vb-\Vb(t)\right)\delta\left(\xb-\Xb(t)\right)}
=
\ensavg{\delta\left(\vb-\ub\left(\Xb(t)\right)\right)\delta\left(\xb-\Xb(t)\right)}
=\\&=
\delta\left(\vb-\ub\left(\xb\right)\right)\ensavg{\delta\left(\xb-\Xb(t)\right)}
=
\delta\left(\vb-\ub\left(\xb\right)\right)\c\left(\xb;t\right)
\end{split}
\ee
Similarly, one can define directly the marginal velocity PDF $\F$ as
\be{vel_pdf_def}
\begin{split}
\F(\vb,t)&
=
\ensavg{\delta(\vb-\ub(\Xb(t)))}=\int\dif\xb\ensavg{\delta(\vb-\ub(\xb))\delta(\xb-\Xb(t))}
=\\&=
\int\dif\xb\,\delta(\vb-\ub(\xb))\ensavg{\delta(\xb-\Xb(t))}=\int\dif\xb\,\delta(\vb-\ub(\xb))c(\xb,t)
=\\&=
\int\dif\xb\,|\deform|^{-1}\delta(\xb-\xb_{0}(\vb))\c(\xb,t)
\end{split}
\ee
where the last step is well defined only locally or when flow field is invertible and $\xb_{0}(\vb)$ is the inverse of the function $\ub(\xb)$ (i.e. the point in space whose velocity is $\vb$). This highlights the relation between the joint PDF $\f$ and the usual spatial concentration PDF $c(\xb,t)=\ensavg{\delta(\xb-\Xb(t))}$ and the role of the velocity gradient $\deform$. In particular, in an homogeneous state, when $c(\xb)=const$, the velocity PDF is equivalent to the inverse of the velocity gradient determinant composed with the inverse velocity function $\xb_{0}(\vb)$
$$
\F(\vb,t) = |\deform|^{-1} ({\xb_{0}(\vb)})
%= \left(|\deform|^{-1} \frobenius \xb_{0}\right) (\vb)
$$

%
%%%%%%%%%%%%%%%%%%%%%%%%%%%%%%%%
%%%%%%%%%%%%%%%%%%%%%%%%%%%%%%%%
%\section{Extensions}
%%%%%%%%%%%%%%%%%%%%%%%%%%%%%%%%
%%%%%%%%%%%%%%%%%%%%%%%%%%%%%%%%
%
%
%%%%%%%%%%%%%%%%%%%%%%%%%%%%%%%%
%\subsection{Relation with CTRW}
%%%%%%%%%%%%%%%%%%%%%%%%%%%%%%%%
%The velocity-marginal PDF of the joint PDF (in the case with or without diffusion) should be equivalent to the Boltzmann-type equation derived by Marco for the CTRW, under some assumptions (yet to find).
%
%To achieve that and derive it from these equations, we could try with
%\begin{itemize}
%\item \sout{Convolution in time to reproduce the sampling done in CTRW}
%\item Approximation of the mixing term \eq{joint_mixing} with a model that spans on the velocity space . 
%The IEM could be seen as an interaction only with the average value while, like in mixing models, more complex (binary) interactions between velocities (concentrations) may be used
%\end{itemize}
%%I think a filtering (convolution) on \eq{fp_joint2} in time might do the job
%%$$\fluct{\f}(t)=\mathcal{F_{\tau}}*\f(t)=\int_{t-\tau}^{t+\tau}\f(s)\mathcal{F}(s-t)\dif s$$
%%after writing $\gradv\f=\frac{1}{\dif\vb}\f(\vb+\dif\vb)-\f(\vb)$
%

\end{document}